# Time evolution of $CO_2$ ro-vibrational excitation in a nanosecond discharge measured with quantum cascade laser absorption spectroscopy

Yanjun Du, Tsanko V. Tsankov, Dirk Luggenhölscher and Uwe Czarnetzki
Institute for Plasma and Atomic Physics, Ruhr University Bochum, D-44780 Bochum, Germany
Email: duyanjun13@gmail.com

**Abstract**

$CO_2$ dissociation stimulated by vibrational excitation in non-equilibrium discharges has drawn lots of attention. Ns-discharges are known for their highly non-equilibrium conditions. It is therefore of interest to investigate the $CO_2$ excitation in such discharges. In this paper, we demonstrate the ability for monitoring the time evolution of $CO_2$ ro-vibrational excitation with a well-selected wavelength window around 2289.0 cm$^{-1}$ and a single CW quantum cascade laser (QCL) with both high accuracy and temporal resolution. The rotational and vibrational temperatures for both the symmetric and the asymmetric modes of $CO_2$ in the afterglow of $CO_2$ + He ns-discharge were measured with a temporal resolution of 1.5 $\mu$s. The non-thermal feature and the preferential excitation of the asymmetric stretch mode of $CO_2$ were experimentally observed, with a peak temperature of $T_{v3,\,max}$ = 966 ± 1.5 K, $T_{v12,\,max}$ = 438.4 ± 1.2 K and $T_{rot}$ = 334.6 ± 0.6 K reached at 3 $\mu$s after the nanosecond pulse. In the following relaxation process, an exponential decay with a time constant of 69 $\mu$s was observed for the asymmetric stretch (001) state, consistent with the dominant deexcitation mechanism due to VT transfer with He and deexcitation on the wall. Furthermore, a synchronous oscillation of the gas temperature and the total pressure was also observed and can be explained by a two-line thermometry and adiabatic process. The period of the oscillation and its dependence on the gas components is consistent with a standing acoustic wave excited by the ns-discharge.

**1. Introduction**

In the last decades, $CO_2$ conversion is of growing interest in the context of greenhouse gas abatement and renewable energy exploration. The non-thermal plasma is a promising means for efficient conversion since the unique electron, vibrational, rotational, and gas temperatures in these plasmas allow focusing the discharge energy to the desired channels instead of heating the gas [1, 2]. Specifically, the vibrational excitation, starting with electron-impact-excitation of lower vibrational levels, followed by vibrational-vibrational (V-V) excitation close to the dissociation threshold level, is a more efficient dissociation pathway [3, 4]. This ladder-climbing population to higher vibrational levels occurs owing to the small difference (~ 0.3 eV) between the higher vibrational levels and faster vibrational-vibrational (V-V) relaxation rates compared to vibrational-translational (V-T) transfer. The electron energy required for vibrational dissociation, which equals the O=CO bond energy (5.5 eV), is much lower than that of the direct electron-impact dissociation (> 7 eV).

However, the detailed physical mechanisms of the vibrational excitation stimulated dissociation is fairly complex. Related questions involve the excitation mechanism by electrons, including interaction of electrons with excited states by superelastic collisions, excitation transfer within one species, e.g. $CO_2$ with $CO_2$, and excitation transfer between different molecules, e.g. $N_2$ with $CO_2$. Lots of efforts have been made to investigate this process both numerically [5-8] and experimentally [9-12]. The temporally resolved measurement of $CO_2$ ro-vibrational excitation is of great importance to gain insight into the excitation and relaxation processes and also the validation of the detailed kinetic modeling of $CO_2$ dissociation.

Different optical diagnostic methods have been used to investigate the $CO_2$ excitation/relaxation process in various types of discharges. Compared with the Raman scattering [11-13] and IR emission[14, 15], the former gives us limited information about the Raman active symmetric bending and stretching modes and the latter presents only qualitative information due to the strong self-absorption, IR absorption spectroscopy is a powerful and preferential method for this topic. Particularly, the broadband IR light source in combination with the Fourier transform infrared spectroscopy (FTIR) provides a relatively large scan range, and thus ensures a simultaneous detection of multiple vibrational modes and even multiple species[16-18]. Rivallan et al. [16] has demonstrated the possibility of detecting the temporally resolved $CO_2$ absorption spectrum with step-scan acquisition mode of FTIR in a $CO_2$/air glow discharge. Klarenaar et al. [17] further developed a fitting approach to deduce the multiple vibrational temperatures and $CO_2$ number density from the broadband absorption spectrum from FTIR. Being able to provide a whole picture of the $CO_2$ excitations in all three vibrational modes, time-resolved FTIR has been widely applied to investigate $CO_2$ dissociation in the pulsed DC glow discharge [18] and RF discharge[19-21].

Despite all these attractive advantages, the accuracy and sensitivity to vibrational temperatures of FTIR are still limited due to the large instrumental broadening. Specifically, since the detected absorbance is a convolution of the real absorbance with large instrumental broadening, the detected absorbance from FTIR is in fact spectrally "smoothed" and averaged, which results in the following effects: (1) the real absorption is generally too strong, even saturated, making the detected absorbance less sensitive to the gas properties (temperature and concentration) considering the exponential relationship in the Beer-Lambert Law. (2) the small absorption peaks from higher vibrational states that are essential for the vibrational temperature determination are always submerged in the final smoothed spectrum, making this method not sensitive to the important vibrational temperature even at low pressure.

The recently developed quantum cascade laser (QCL), especially the fast frequency chirp of pulsed QCLs, known as the *intrapulse* mode [22], makes it possible to probe the time-resolved absorption spectrum[23-25]. More recently, Damen et al. [26, 27] proposed to measure the time evolution of $CO_2$ excitation with the intermittent scanning of three QCLs operated at a pulsed mode with a temporal resolution ~100 $\mu$s. With decent efforts to avoid the rapid passage effects [28, 29] and jitter [30], which are the inherent defects of this *intrapulse* operation mode, improved accuracy and sensitivity to vibrational temperature were achieved compared with FTIR. However, further improvement in the temporal resolution is challenging due to the compromise between the total scan range and the ultimate temporal resolution (laser pulse length). It is generally limited to the order of tens of $\mu$s estimated with a downchirp rate of several MHz/ns and a scan range of ~1 $cm^{-1}$ to cover enough transitions from different vibrational modes of $CO_2$, although it varies with lasers. In addition, the measurement accuracy of absorbance is strongly subject to the pulse to pulse fluctuations in both laser intensity and chirp rate[25].

In this work, the nanosecond pulsed discharge, which enables a good separation and control of the electron excitation and the collisional ro-vibrational excitation, will be used to investigate the $CO_2$ vibrational excitation. To achieve both high accuracy and temporal resolution, a new method is demonstrated with a well-selected wavelength window and a single QCL operating in a continuous slow scan (or step scan) mode. With this method, rotational and multiple vibrational temperatures together with the $CO_2$ density are simultaneously determined with a temporal resolution of 1.5 $\mu$s and high accuracy as well. The fast excitation and relaxation process for all three vibration modes of $CO_2$ in the afterglow of the $CO_2$-He nanosecond pulsed discharge will be presented and discussed.

## 2. Experimental setup and measurement strategy

### 2.1 Nanosecond pulsed DC discharge

The experimental setup is schematically shown in figure 1. The nanosecond discharge used in this paper is similar to that in reference [31]. To be brief, the discharge consists of two molybdenum electrodes with a cross-section of 20 × 1 mm placed vertically with a distance of 1 mm. Two glass plates are pressed tightly to the electrodes in front and backward, yielding a well-confined discharge volume of 20 mm ×1 mm ×1 mm. Nanosecond voltage pulse generated by combining a DC power supply (LNC 6000-10 neg) with a fast high-voltage switch (Behlke HTS-81) is applied to one electrode with the other one grounded. A delay generator (Stanford Research Systems DG535) is used to trigger and open the switch with a repetition rate of $f_p$ = 2 kHz and an on-time of 150 ns. The applied voltage and current are monitored by a high-voltage probe (Tektronix-P6015A) and current probe (Lecroy AP015) in conjunction with an oscilloscope (Teledyne Lecroy, WaveSurfer 510). The incoming gas consists of 10% $CO_2$+He by mixing pure $CO_2$ and He gas flow using two mass flow controllers (MKS instruments) with a total flow rate of 30 sccm. The gas mixture is guided into the discharge gap through a hole in the center of the backward glass plate and leaves the discharge from both ends, as shown in Fig.1 (b). The pressure in the discharge chamber is monitored by a pressure gauge (Pfeiffer vacuum) and is kept constant at 145 mbar by fine adjusting the mechanical needle valve at the gas outflow. At this condition, the gas residence time is around 6 ms, suggesting that all molecules experience 12 pulses before leaving the discharge. The discharge emission is visually homogeneous within the whole discharge gap.

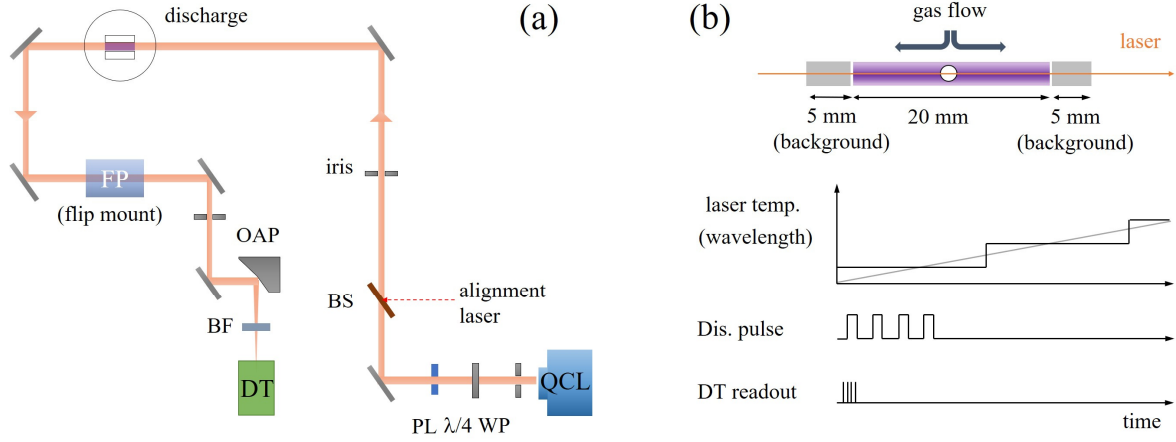

Figure 1. (a) Schematic of the experimental setup, including the discharge chamber and the optical alignment. WP: wave plate, PL: polarizer, BS: beam splitter, OAP: off-axis parabolic mirror, BF: bandpass filter, DT: detector; (b) Schematic view of the laser light path and the trigger scheme for the high temporal resolution measurement.

*2.2 Optical system and measurement strategy*

The laser system to probe $CO_2$ absorption spectrum is rather compact and simple, as shown in Fig.1 (a). A single-mode CW quantum cascade laser (Alpes) with a coverage from 2276 to 2290 cm$^{-1}$ was used to scan the $CO_2$ absorption transitions. The collimated output of the laser is guided in and out of the discharge chamber, through two inserted wedged $CaF_2$ windows. A combination of polarizer and a quarter-waveplate is used to adjust the laser intensity to avoid saturation. Two irises help to prevent the potential etalon effects in the optical pass by shielding the back reflection from the beam splitter and the $CaF_2$ windows. After traversing the discharge chamber, the laser beam is focused onto a DC-coupled photovoltaic detector (Vigo, PVI-3TE-5, bandwidth 1 MHz) by an off-axis parabolic mirror. A narrow bandpass filter (Thorlabs, FB4250-500) and a third iris are mounted in front of the detector to reduce the infrared thermal emission from the discharge. The laser frequency is characterized by a silicon etalon (LightMachinery) with a free spectral range (FSR) of 0.0176 cm$^{-1}$ mounted on a flipping holder.

High temporal resolution is achieved by operating the QCL with a slow scan, which can be regarded as quasi-step-scan mode, as shown in the trigger scheme in Fig. 1(b). A ramp signal generated from a function generator (Agilent 33250A) is sent to a commercial laser controller (ILX 3736) to scan the temperature of the QCL with a frequency of $f_l$ = 10 mHz and an amplitude of 5°C (~0.2 cm$^{-1}$/°C). The temperature tuning is adopted as it provides a wider wavelength sweeping range and a smaller intensity variation compared with the current scan. The latter is crucial considering the generally limited dynamic range of the fast IR detector. With such a slow temperature scan, the laser wavelength can be regarded as constant within 20 ms (5 °C / 100 s × 0.02 s = 0.001°C), considering the minimal temperature tuning step of the controller (0.001°C). For each wavelength, the temporal profile of 40 ($f_p$ = 2 kHz) repeated discharge pulses are acquired with the detector. The light intensity from the detector is read out by an oscilloscope (PicoScope 5100D) with a vertical resolution of up to 16-bit. With this measurement strategy, the highest temporal resolution is generally limited by the sampling rate of the oscilloscope and the bandwidth of the detector. Although the limited bandwidth of the detector makes it impossible to describe the rapid discharge phase, achieved the temporal resolution of 1.5 μs is high enough to describe the excitation/ deexcitation process in the afterglow (in the order of 100 μs). A measurement series is performed in the following procedure:

(1) Measure the incident laser intensity $I_0$ by purging the chamber with pure He at 145 mbar;

(2) Measure the wavelength tuning property by inserting the etalon;

(3) Measure the transmitted light $I_t$ with the desired $CO_2$/He mixture at 145 mbar with discharge turned on.

With all three spectra, the spectrally and temporally resolved $CO_2$ absorption can be processed and reconstructed. Since each spectrum takes 100 s, the total acquisition time for one measurement is around 5 min. Additionally, one more spectrum measured with the desired gas mixture before turning on the discharge could help to get a self-calibration for the $CO_2$ concentration and collisional broadening coefficient, but it is not a must for each measurement. Also, it should be noted that since the laser intensity and frequency are measured individually, the temperature tuning signal is always monitored as a synchronization.

## 3. Theoretical analysis

### 3.1 Calculation of $CO_2$ absorption spectrum under non-equilibrium condition

Unlike the conventional tunable diode laser absorption spectroscopy (TDLAS), which uses the integrated areas or broadenings of two individual lines to infer gas temperature and species concentration, fitting a broadband absorption spectrum with several transitions is necessary in this paper to obtain the multiple temperatures ($T_{rot}$, $T_{v1}$, $T_{v2}$, and $T_{v3}$) and concentration of $CO_2$. For the fitting this spectrum first needs to be calculated as a function of the various temperatures.

The calculation algorism is implemented based on the HITEMP-2010 database [32], which contains the line-by-line information of $CO_2$ for 12 isotopes and hot bands up to $v_3 = 6$. The notation of a vibrational state in this paper follows the Herzberg rule [33] with an extra ranking index $r$ to classify the Fermi resonating group ($v_1$ $v_2^{l_2}$ $v_3$ $r$). ($r = 1, 2, 3, ..., v_1+1$; $r = 1$ denotes the highest vibrational level within a resonating group). In this terminology system there only exist transitions with $l_2 = v_2$ since all other levels ($v_1$ $v_2^{l_2}$ $v_3$) ($l_2 \neq v_2$) are resonated with (($v_1$-$l_2$/2) $l_2^{l_2}$ $v_3$) levels, although theoretically $l_2$, representing the contribution of bending mode to the angular rotation, can take values from $v_2$, $v_2$-2.... The absorbance $A(v)$ or fractional absorption (used when the absorption is too strong or even saturated), as described by the basic Beer-Lambert Law, can be calculated as follows,

$$A(v) = -\ln(I_t(v)/I_0(v)) = S(T) pxl\sigma(v), \quad (1)$$

where $I_t(v)$ and $I_0(v)$ are the transmitted and incident light intensities, respectively. $p$ and $l$ are the total pressure [atm] and optical length [cm], determined by the experimental conditions. $x$ is the mole fraction of the $CO_2$ and $\sigma(v)$ is the line profile.

The linestrength $S(T)$ [cm$^{-1}$(atm×cm)$^{-1}$] at a certain temperature can be calculated from the Einstein coefficients from HITEMP as shown in [17, 27] with special attention paid to the definition of all derived parameters [34]. A better and easier way of calculating the linestrength at a specific temperature, or temperature set $T = [T_{v1}, T_{v2}, T_{v3}, T_{rot,}]$ under non-equilibrium condition, is the rescaling method,

$$S(T) = S(T_{ref}) \cdot \frac{f''(T)}{f''(T_{ref})} \cdot \left[ \frac{1 - e^{-\Delta E/k_B T}}{1 - e^{-\Delta E/k_B T_{ref}}} \right], \quad (2)$$

where $f''(T)$ is the fraction of population in the lower state at a specific temperature, $\Delta E$ is the energy difference of the upper and lower state and can be obtained from HITEMP database. $S(T_{ref})$ is the standard linestrength listed in HITEMP at a reference temperature of $T_{ref} = 296$ K. In this approach, all the inherent constants associated with a specific transition are canceled out in the rescaling method. Only the terms that represent the rescaling factor due to the number density (second term in Eq. (2)) and stimulated emission (last term) at the target and reference temperature still remain. Therefore, the calculation of the $CO_2$ absorption spectrum is simplified to the calculation of the number density $f''(T)$.

To calculate the fraction of population $f''$ under non-equilibrium condition, the total energy of a specific level ($v_1$ $v_2^{l_2}$ $v_3$ $r$ $J$) should be partitioned into different vibrational and rotational modes and scaled with corresponding temperatures,

$$f'' = f_{v1} \cdot f_{v2} \cdot f_{v3} \cdot f_J \quad (3)$$

where $f_J$ and $f_{vi}$ are fractions of population in the pure rotational and vibrational dimensions. Considering the fast translation-rotation processes (T-R) process (with a characteristic time in the order of tens of ns under our conditions), the rotational temperature is in equilibrium with the gas temperature $T_g$ and the rotational distribution function always follows the Boltzmann distribution. The vibrational distribution can be well described by a Treanor distribution especially for the $v_3$ mode, which has a faster V-V exchange rate than that of V-T. Therefore,

$$f_{vi} = \frac{g_{vib,i}}{Q_{vib,i}} \exp\left(-c_2 \left(\frac{E_{h,i}}{T_{vib,i}} + \frac{E_{a,i}}{T_{rot}}\right)\right)$$

$$f_J = \frac{g_i \cdot g_s \cdot (2J+1)}{Q_{rot}} \exp\left(-c_2 \cdot \frac{E_{rot,J}}{T_{rot}}\right) \quad (4)$$

where $g_{vib,i}$ is the degeneracy of each vibrational mode and equals 1 for the $v_1$ and $v_3$ modes. For the doubly degenerate bending mode $v_2$, since the vibrations in two orthogonal planes have the same vibrational energy and any linear combination of the two eigenfunctions is also an eigenfunction of the same energy level, the degeneracy is $v_2 + 1$. $E_{h,i} = v_i\omega_i$, and $E_{a,i} = -v_i(v_i-1)\omega_e x_e$ are the harmonic and anharmonic components of the vibrational energy, which are scaled with the vibrational temperature $T_{vi}$ and rotational temperature $T_{rot}$ in the Treanor distribution[35], respectively. $\omega_i$ and $\omega_e x_e$ are derived from the vibrational constants from paper [36], similar as in [17, 27]. As for the rotational term, the $g_i$ and $g_s$ are the rotational state-independent and state-dependent nuclear spin statistical weight.

$g_s$ depends on the total spin of the interchanged identical nuclei in a molecule and will cause the vanishing of symmetric (or asymmetric) rotational state and thus should be discussed accordingly for different $CO_2$ isotopologues [34]. The partition sum $Q_{vib,i}$ and $Q_{rot}$ are separately calculated under non-equilibrium condition by numerical summation of all the vibrational and rotational levels with a truncation of $v_1 = 20$, $v_2 = 43$, $v_3 = 13$, and $J = 150$.[37] The truncation is validated by checking the convergency of $Q_{tot} = Q_{rot} \cdot Q_{vib,i}$ up to 3000 K (> 99.93%) and by comparison with the tabulated database from [38] under equilibrium condition.

The line profile $\sigma(v)$ [cm$^{-1}$] is described by a Voigt profile with consideration of the line-dependent Doppler and collisional broadenings. To accelerate the simulation and thus the fitting algorism, the direct summation formed Voigt profile, proposed by Liu et al. [39] is used with good accuracy. At our experimental conditions ($T_g$ close to room temperature), the half-width at half-maximum (HWHM) of the Doppler broadening determined by the gas temperature, $\delta_G$, is around 0.002 cm$^{-1}$ at our experimental condition and is one order of magnitude smaller than that of the collisional broadening calculated as follows,

$$\delta_c(P,T) = p \cdot \left[ \left(\frac{T_{ref}}{T}\right)^{n_{CO_2}} \cdot \Upsilon_{CO_2} x_{CO_2} + \left(\frac{T_{ref}}{T}\right)^{n_{He}} \cdot \Upsilon_{He}(1 - x_{CO_2}) \right], (5)$$

where $\Upsilon_{CO_2}$ and $\Upsilon_{He}$ are the self-broadening and He-broadening coefficient of $CO_2$ at reference condition ($T_{ref}$ = 296 K and $p_{ref}$ = 1 atm) and $n_{CO_2}$ and $n_{He}$ are the corresponding temperature exponents. Due to the lack of detailed line-by-line data, $n_{CO_2}$ and $n_{He}$ are assumed to be equal to $n_{air}$ from HITEMP database, which is reasonable considering the low gas temperature in our experimental condition. The collisional broadening coefficient $\Upsilon_{He}$, is calculated based on the tabulated data $\gamma_{air}$ from HITEMP with $\Upsilon_{He} = q \cdot \Upsilon_{air}$ assuming the same dependence on rotational quantum $J$ but a global quenching factor to correct the difference between He and air. The quenching factor $q$ will be measured and discussed later. The collisional broadenings due to the potential dissociation products $O_2$ or CO are ignored in our experiments considering the low dissociation rate and low $CO_2$ concentration.

*3.2 Validation of simulation under both equilibrium and non-equilibrium conditions*

The code simulating the absorption spectrum is validated by comparisons with the equilibrium $CO_2$ absorbance calculated with HAPI [40] (HITRAN python interface to calculate equilibrium spectra) and measured non-equilibrium absorption spectrum from the literature [17, 41].

Fig. 2 shows a comparison of the equilibrium absorbance for pure $CO_2$ around 2289.5 cm$^{-1}$ calculated from HAPI with $T$ = 1000 K and the established non-equilibrium simulation code with $T_{rot} = T_{v1} = T_{v2} = T_{v3} = 1000$ K. The moderate pressure, $p = 0.03$ atm, is chosen to ensure a comparable contribution of the Gaussian and the Lorentzian broadening, thus the broadening calculation and the line profile could also be inspected. Meanwhile, since the total energy and the total internal partition sums used in HAPI are calculated ab initio [38], the perfect agreement between these two simulations also suggests the accurate calculations of the total energy and of the partition function in the rescaling model. It should also be noted that when calculating the equilibrium spectra, the energies of the Fermi-resonant levels are additionally corrected by calculating the Fermi coupling off-diagonal elements as shown in [36, 42].

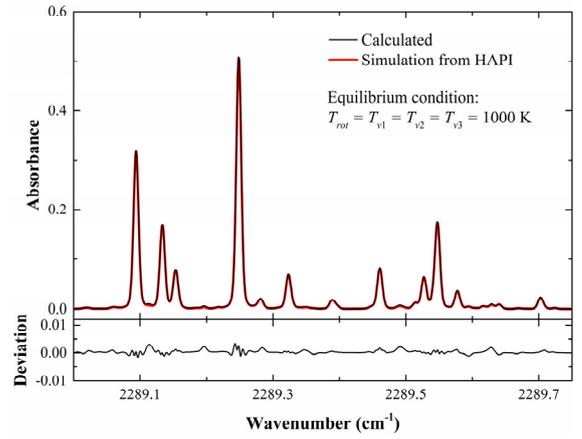

Figure 2. Comparison of the calculated absorption with HAPI at equilibrium condition ($T_{rot} = T_{v1} = T_{v2} = T_{v3} = 1000$ K) for 10% $CO_2$, $p = 0.03$ atm and $l = 1$ cm.

To further validate the splitting of the total energy of a specific energy level into different rotational and vibrational modes in the proposed simulation code, the experimental transmittance under non-equilibrium condition from the literature [41] is re-investigated and fitted with our fitting codes. Fig. 3 shows the measured fractional absorption spectrum from [41] with a mixture of 10% $CO_2$, 38% $N_2$, and 52% He at 15 Torr around 2284.4 cm$^{-1}$. Owing to the relatively low pressure, the detected absorption peaks are well separated. The corresponding dominant transitions are identified and labeled to indicate the dependence of the peak intensities on the different vibrational and rotational temperatures. Two different fits are made. For the first one (blue curve in Fig. 3) $T_{v1} = T_{v2}$ is assumed, whereas for the second one (black curve) $T_{v1}$ and $T_{v2}$ are considered independent. As can be seen, both fitting results agree very well with the experimental results. The values for the temperatures obtained from the fits are: $T_{rot} = 495 \pm 9$ K, $T_{v1} =$ $T_{v2}$ = 508 ± 7 K, $T_{v3}$ = 2634 ± 60 K for the first case and $T_{rot}$ = 498 ± 9 K, $T_{v1}$ = 522 ± 10 K $T_{v2}$ = 500 ± 8 K, $T_{v3}$ = 2658 ± 63 K for the second case.

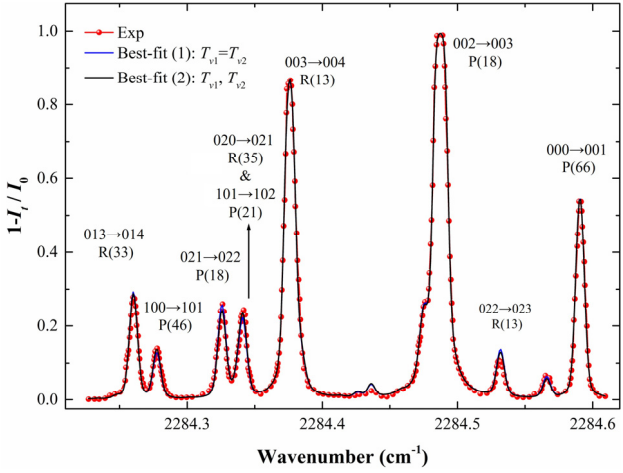

Figure 3. Fitting of a non-equilibrium $CO_2$ absorption from the literature [41]. Each peak is denoted as ($v_1$ $v_2$ $v_3$) with $l$ and $r$ omitted to reduce the complexity of the notation.

Several conclusions can be made based on these fitting results. Firstly, the approximate equality of fitted vibrational temperatures $T_{v1}$, $T_{v2}$ for the second case and $T_{v12}$ in case (1) corresponds well with the conclusion obtained from a Boltzmann plot in [41]. This confirms the assumption used in the following discussion that $T_{v1}$ = $T_{v2}$ due to the strong coupling between the symmetric stretch and bending modes, i.e. Fermi resonance[17, 43]. Secondly, although no detailed temperatures were provided for this example transmittance in [41], the fitted temperatures match well with those from [17, 27, 44] with a similar fitting algorithm, which again verifies the accuracy of the simulation code and partition of rovibrational energy. Thirdly, it is also worth noticing that the relatively large fitting uncertainty (calculated with 95% confidence bounds), is attributed to the non-ideal selection of wavelength range, i.e. insufficient energy gap, and the almost saturated strong absorptions, both of which reduce the sensitivity of the detected absorbance to the gas temperature or, equivalently, the concentration. This is also one of the reasons why in addition to this example absorbance, several other wavelength intervals, from 2140 to 2310 cm$^{-1}$, were needed to be scanned in [41] to obtain accurate rotational and vibrational temperatures. Therefore, a well-selected wavelength is essential for the accuracy of the results when the total scan range is limited.

### 3.3 Transition (wavelength) selection and fitting strategy

Considering the limited scan range of a single-mode MIR laser, a well-selected wavelength range is needed for the simultaneous determination of multiple temperatures and concentration. Except for a reasonable absorption strength, which should be neither too small to ensure a good SNR, nor so big to cause an absorption saturation and thus not sensitive to the fitting parameters, some other nontrivial rules need to be taken into account while selecting the wavelength window. Since there are at least five parameters $T_{rot}$, $T_{v1}$, $T_{v2}$, $T_{v3}$ and $n_{CO2}$ to be determined, at least five independent equations or constraints are required to avoid an underdetermined system. The following guidelines were used when selecting the wavelength range in this work.

(1) To obtain accurate $T_g$ or $T_{rot}$: one pair of lines from the same vibrational level (better to be the ground vibrational level) with a difference in rotational level $J$ as large as possible should be included;

(2) To obtain accurate $T_v$s: at least one transition from each vibrational mode, i.e. symmetric stretching ($v_1$x$^x$xx,), symmetric bending (x$v_2^x$xx), and asymmetric stretching (xx$^x v_3$x) with $v_1$, $v_2$, $v_3$ > 0 should be included;

(3) To increase the reliability of the individual vibrational temperatures, transitions with multiple quanta excited, e.g. $v_1 v_2^x$xx, $v_1$ & $v_2$ >0 are recommended.

Table 1. Spectroscopic parameters for the selected $CO_2$ transitions from HITEMP 2010 [32]

| Label | Frequency (cm$^{-1}$) | Vib. | Rot. | Energy (cm$^{-1}$) |
|---|---|---|---|---|
| 0 | 2288.784 | 00$^0$01 | R6 | 16.39 |
| 1 | 2289.094 | 01$^1$01 | P51 | 1702.40 |
| 2 | 2289.134 | 02$^2$01 | P39 | 1945.79 |
| 3 | 2289.249 | 00$^0$01 | P62 | 1522.16 |
| 4 | 2289.396 | 01$^1$01 | R24 | 883.18 |
| 5 | 2289.548 | 10$^0$02* | P42 | 1990.11 |
| a | 2289.323 | 03$^3$01 | P26 | 2278.63 |
| b | 2288.671 | 11$^1$01 | P29 | 2416.42 |
| c | 2288.476 | 10$^0$01 | P42 | 2092.49 |
| A | 2289.461 | 00$^0$11 | P39 | 2952.76 |
| B | 2289.578 | 01$^1$11 | P26 | 3276.04 |
| C | 2288.601 | 01$^1$11 | P27 | 3297.41 |

* Notation with consideration of Fermi-resonance. Theoretically, it belongs to the vibrational level (02$^0$0) as shown in the energy level scheme in Fig. 4(a).

Based on the abovementioned guidelines, a wavelength window of ~1 cm$^{-1}$ near 2289 cm$^{-1}$ is chosen with all noticeable transitions in that range listed in Table 1. The transitions are split into different groups based on their distinct characteristics. For example, all transitions labeled by a number from "0~5" are the lines that have a measurable absorption at room temperature.

These lines can be detected without a discharge as a benchmark of the wavelength window. The transitions that are sensitive to the vibrational temperature of the symmetric ($T_{v1}$ or $T_{v2}$) and the asymmetric modes ($T_{v3}$) are marked by lowercase and uppercase alphabetic characters, respectively.

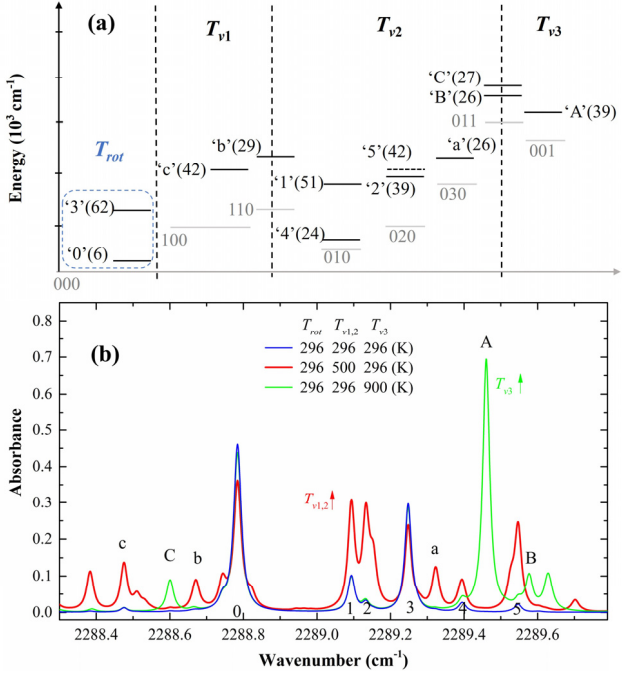

Figure 4. (a) Simplified energy level scheme for the transitions used in this paper and the corresponding sensitivity to different vibrational and rotational temperatures. The label of each peak is denoted as ($v_1$ $v_2$ $v_3$) with $l$ and $r$ omitted to reduce complexity. (b) The simulated absorption spectra in the selected wavelength range at room temperature and at non-equilibrium conditions with $P = 145$ mbar, 10% $CO_2$, $l = 2$ cm.

To further demonstrate the high sensitivities of the selected transitions to different temperatures, a simplified energy level scheme for the lower ro-vibrational states of the selected transitions are plotted in Fig. 4(a) with the respective vibrational state marked as a grey line below and the rotational quantum "$J$" in the bracket. Peaks "0" and "3", both belonging to transitions to the ground state, are a perfect line pair for the rotational temperature measurement due to their large $J$ difference up to 56 and comparable intensities (see Fig. 4(b)). Furthermore, in addition to the coverage of the transitions from all three vibrational modes, the selected wavelength window also includes absorption lines that have a lower level with multiple vibrational modes excited, e.g. "b", "B" and "C". Therefore, the population density of these levels depends on a combination of the vibrational temperatures. Then the fitting quality of these peaks is a measure of the reliability of the determination of the individual vibrational temperatures. Fig. 4(b) compares also the calculated absorption in the target wavelength range under our experimental condition (10% $CO_2$, $P = 145$ mbar, $l = 2$ cm) at room temperature and at elevated vibrational temperatures. As can be seen, in addition to a reasonably strong absorption intensity for all peaks, the strong sensitivity of the target transitions to the different temperatures is clearly observed. Consequently, this wavelength window provides good sensitivity to the various parameters of interest.

## 4. Results and discussions

Figure 5 shows a representative example of the detected spectra required for a time series measurement. The sync signal from the laser temperature tuning, as shown in yellow, is used to synchronize all spectra in time although they are all measured individually. As shown by the etalon fringes, the QCL is tuned over 1.14 cm$^{-1}$ ( = 65 fringes $\times$ 0.0176 cm$^{-1}$) within 50 s to probe all the transitions under investigation. To obtain the absolute frequency, firstly the signal from the etalon is averaged over a period of 20 ms. Then the positions of the fringe peaks are read out and fitted by a third-order polynomial. The absolute wavelength is obtained by using the position of the absorption peak at 2288.78 cm$^{-1}$.

Beside the three spectra required for a time series measurement, also the spectrum recorded with the desired gas mixture before turning on the discharge is included, in Fig. 5. Compared with the smooth incident light intensity, $I_0$, absorption peaks that have detectable intensity at room temperature and marked as 0-5 in section 3.3 are observed with the $CO_2$ gas mixture as $I_t$. This spectrum allows an experimental estimation of the quenching factor $q$, ensures no systematic error in the gas mixture, and also provides a closer inspection of the stability of the whole system. The transmitted light intensity with discharge is shown in red. Prominent differences are observed in comparison to the one without a discharge. As can be seen, transitions from high vibrational levels, marked as alphabetic characters in section 3.3 appear when the discharge in turned on. The recorded spectra then provide a temporally resolved measurement of the $CO_2$ absorption in the afterglow of ns-discharge can be obtained and further discussed.

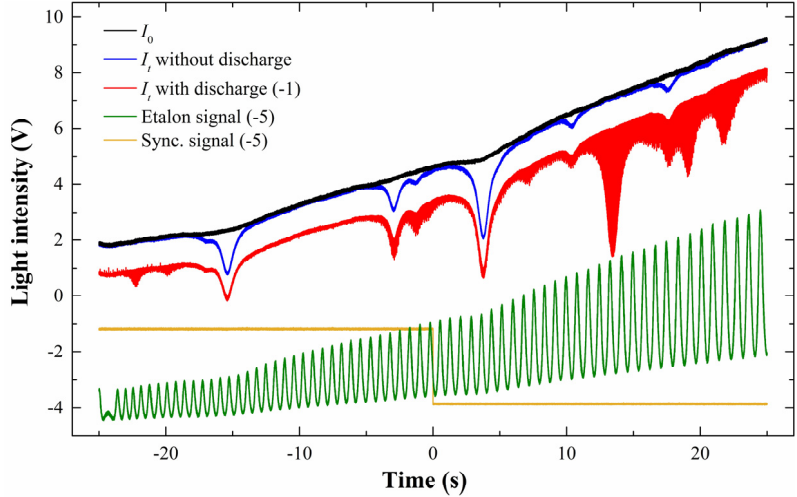

Figure 5. Example of the detected spectra for a time series measurement. The QCL is scanned by temperature with $f$ = 10 mHz. The discharge is operated with 10% $CO_2$ + He at 2 kHz with a voltage of 3 kV, current of ~ 6 A, and a pulse length of 150 ns. The light intensities of $I_t$ with discharge and the etalon signal are shifted vertically for clarity.

### 4.1 Measurement of collisional broadening coefficient $\gamma_{He-CO2}$ under equilibrium conditions

Since the collisional broadening of $CO_2$ is dominant at our experimental pressure and the 90% collisional partner is He, it is important to have an accurate collisional coefficient of $CO_2$ with He, $\gamma_{He}$. The arbitrary assumption that $\gamma_{He}$ equal to $\gamma_{air}$ listed from HITRAN or HITEMP database due to the lack of detailed data from literature, will result in an obvious discrepancy in the best-fit spectrum and an overestimation of the final fitting $CO_2$ concentration due to the overestimation of the broadening effect. Here, we conducted an experimental measurement of $q$ by checking the $CO_2$ absorption spectrum under equilibrium conditions.

Fig. 6 shows the measured and best fitted $CO_2$ absorbance within the target range under equilibrium conditions. The red circles are the measured spectrum with 10% $CO_2$ + He at a pressure of 145 mbar without discharge. The best-fit spectrum with collisional coefficient $q$ as an extra fitting parameter is shown in black. As can be seen, perfect agreements are observed for all peaks and the best fit parameters are as follows, $T_{rot}$ ( = $T_{v1} = T_{v2} = T_{v3}$) = 297.5 ± 0.2 K, $n_{CO2}$ = 9.8 ± 0.1 % and $q$ = 0.80 ± 0.01. The best-fit gas temperature $T_{rot}$ = 297.5 ± 0.2 K is a reasonable room temperature in the lab and the best-fit $CO_2$ concentration also agrees well within uncertainty with the preset value. In addition, this overall quenching factor $q$ = 0.80 ± 0.01 obtained from the fitting yields a collisional broadening coefficient $\gamma_{He}$ = 0.062 cm$^{-1}$/atm and 0.055 cm$^{-1}$/atm for the strong absorption peak R(6) for R(62), respectively. These values correspond well with those from the literature [45, 46]. The collisional broadening coefficients of $CO_2$ with He for other target transitions, which are not observed at room temperature, are all rescaled with this $q$ factor, assuming the same rotational quantum $J$ dependence as with air.

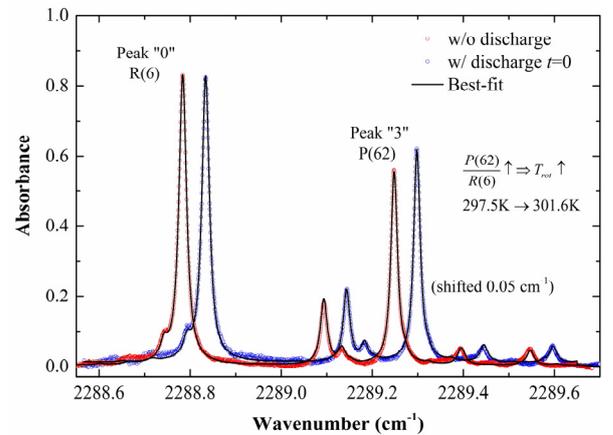

Figure 6. Absorbance of the $CO_2$ gas mixture without a discharge (red symbols) and before the discharge pulse ($t$ = 0 $\mu$s) (blue symbols) for the target scan range. The blue curve is shifted by 0.05 cm$^{-1}$ horizontally for clear comparison.

The detected $CO_2$ absorption spectrum and its best-fitting with the same gas mixture but at a slightly elevated gas temperature under equilibrium is also included in Fig. 6. This spectrum is obtained while running the discharge in a single-pulse mode by increasing the gas flow rate to 400 sccm, thus the light path is filled with the pre-heated fresh gas mixture before the discharge pulse at $t$ = 0. As expected, the best-fit gas temperature and $CO_2$ concentration are 301.6 ± 0.2 K and $n_{CO2}$ = 9.8 ± 0.1 %. As can be seen in Fig. 6, even with a temperature increase of only 4.1 K, the intensity of peak "3" that belongs to a high rotational state, and thus the ratio of peak intensity between peak "3" and "0", increase notably due to the large energy difference between these two lines. This demonstrates the high accuracy of this method and confirms the good selection of the wavelength window.

In addition to inferring the collisional factor $q$, the stability of the measurement method and optical system can be evaluated with the absorption spectrum measured without discharge. In fact, the absorbance shown in Fig. 6 was reconstructed in the same procedure with the temporally resolved measurement with discharge by recording a hypothetical discharge pulse signal with high voltage off. By fitting the time-resolved "stable" background gas mixture absorbance without discharge, the stability of this method could be accessed. The averaged best-fit rotational temperature, $CO_2$ concentration and quenching factor over 500 $\mu s$ are 297.50 K, 9.794% and 0.799, with a standard deviation of 0.04 K, 0.007% and 0.001, respectively. It is clear that the standard deviation of the background mean value is smaller than the fitting uncertainty within the 95% confidence bound error of the fitting process, which suggests the good stability of the experimental system and the method over time.

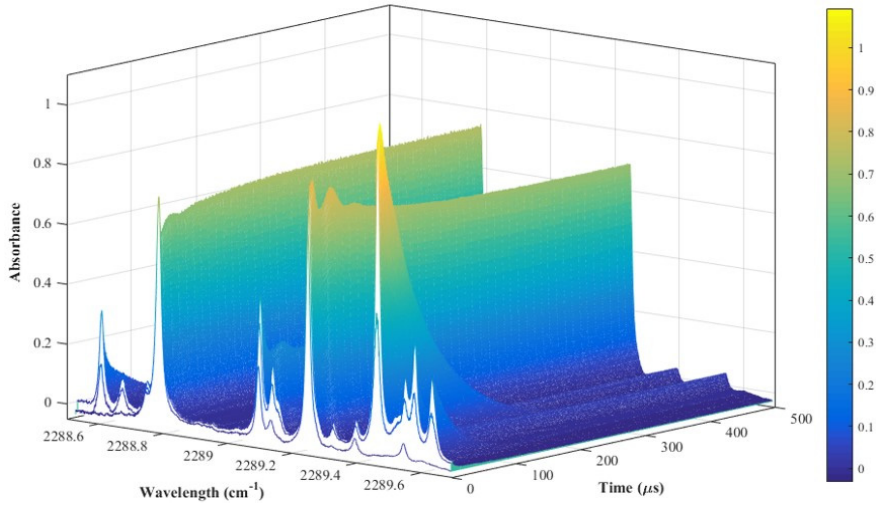

Figure 7. Time evolution of the measured absorbance in the afterglow of the nanosecond discharge. The discharge is operated with 10% $CO_2$ + He at 145 mbar with a frequency of 2 kHz, a voltage of 3 kV ($I \sim 6A$), and a pulse length of 150 ns.

*4.2 Temporally resolved measurement*

Fig. 7 presents the temporally and spectrally resolved absorption of $CO_2$ in the nanosecond $CO_2$-He discharge at 145 mbar with a frequency of 2 kHz, voltage of 3 kV and a pulse length of 150 ns. As can be seen, immediately after the discharge pulse, the peaks from the excited vibrational states appear and rise dramatically within the first 3 $\mu s$, which suggests a distinct vibrational excitation of $CO_2$ due to the discharge. After that, a clear decay is observed for all these peaks. In contrast, the change in peaks 0 and 3 from the ground vibrational state is relatively moderate during the whole process, except for an oscillation within the first 100 $\mu s$, indicating a mild variation in the rotational temperature. Through the fitting of the temporally resolved absorption spectra, the evolution of the $CO_2$ rotational and vibrational temperatures can be obtained.

Several representative absorption spectra at different time points are shown in Fig. 8, together with the corresponding best-fitting. At $t = 0$ $\mu s$ just before the discharge pulse, local thermal equilibrium is reached along the whole light pass, including both discharge area and background (the 5 mm gap between discharge and windows, as shown in Fig. 1(b)) since only transitions from the low energy states are observed. The best-fit temperature is $T_{rot} = T_{v1,2} = T_{v3} = 311.1 \pm 0.2$ K, suggesting a negligibly small gas heating in this highly non-thermal ns-discharge. The best-fit $CO_2$ concentration reduces to 9.4 ± 0.1%, indicating a dissociation degree of only 4% of the $CO_2$ gas. This $CO_2$ dissociation degree is close to the value reported from a pin-to-sphere configured ns-discharge [5, 9] with a similar SEI (specific energy input) of 0.86 eV/molecule (calculated with mean pulse energy around 1mJ). Considering the small gas flow rate and the interchange between the discharge area and the background, the temperature and $CO_2$ concentration of the background gas are assumed to be constant in the overall afterglow with those at $t = 0$ $\mu s$. The green line in Fig. 8(b) depicts the spectral contribution of the background gas to the final line-of-sight spectrum. It is worth noting that the background has limited influence on the final fitting results due to its small peak intensities, especially for the determination of vibrational temperatures, although the optical length of background gas is around 50% of the discharge area.

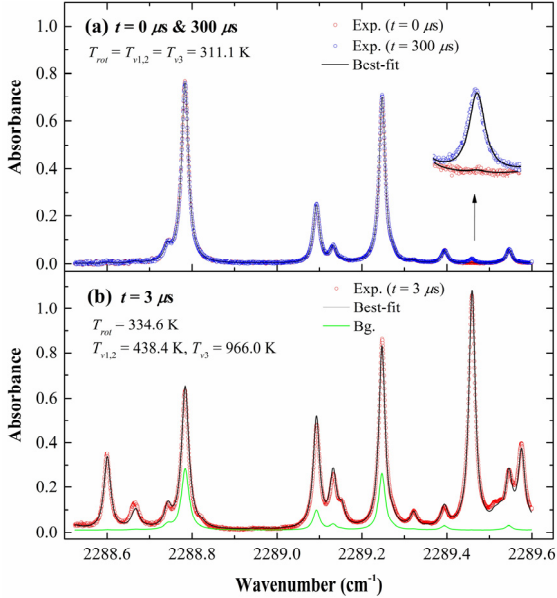

Figure 8. Examples of fitted absorption spectra for 10% $CO_2$ + He with a frequency of 2 kHz, voltage of 3 kV and a pulse length of 150 ns. (a) Before discharge pulse $t = 0$ μs and $t = 300$ μs; (b) $t = 3$ μs when $T_{v3}$ reaches the maximum.

The absorption spectrum of $CO_2$ at $t = 3$ μs, changes dramatically as shown in Fig. 8(b). All three vibration modes are highly excited, especially the asymmetric stretch mode, as shown by the distinct peak "A". The best-fit temperatures are $T_{rot} = 334.6 \pm 0.6$ K, $T_{v1,2} = 438.4 \pm 1.2$ K and $T_{v3} = 966.0 \pm 1.5$ K. It is clear to see excellent agreement exists between the measurement and the best-fit in this strongly non-equilibrium condition. As shown in the temporally resolved absorption profile in Fig. 7, after $t = 3$ μs the intensity of the transitions from excited vibrational states decay fast with time and finally reach zero. An example of the measured absorbance at a later phase $t = 300$ μs is also attached in Fig. 8(a). In comparison with the spectrum at $t = 0$ μs, all transitions overlap with each other except for peak "A" that belongs to the vibrational state of (001). This suggests that until $t = 300$ μs, $T_{rot}$, $T_{v1}$, and $T_{v2}$ have already decayed to the same value as in the equilibrium condition before the discharge pulse, while $T_{v3}$ is still higher than the others. This is also reflected in the best-fit temperatures: $T_{rot} = 311.9 \pm 0.2$ K, $T_{v1,2} = 312.0 \pm 0.4$ K and $T_{v3} = 443 \pm 2$ K, indicating a slower deexcitation of the asymmetric stretch mode of $CO_2$ compared with the symmetric stretch or bending modes, as will be discussed later.

By fitting the overall time-resolved absorbance in Fig. 7, the evolution of $T_{rot}$, $T_{v1,2}$ and $T_{v3}$ in the afterglow of the nanosecond discharge can be obtained, as shown in Fig. 9. It should be noted that during the overall fitting process, no extra assumption or a priori constrains are used in the fitting to ensure a self-consistent result, except for the time-invariant parameters for the background $CO_2$. As can be seen, before the discharge pulse, $t = 0$ μs, thermal equilibrium is obtained among all degrees of freedom, i. e. translation, rotation, and all three vibration modes. During the 150 ns discharge pulse, the $CO_2$ molecules get excited by the energetic electrons. Further effort will be made in the future to look into the discharge phase with a temporal resolution of ~ ns by replacing the detector with the one with a larger bandwidth (> 100 MHz). After the discharge pulse, the vibrational temperatures as well as the gas temperature keep increasing and reach the maximum value at around 3 μs after the pulse. Pronounced excitation of the asymmetric mode is observed with a peak vibrational temperature of $T_{v3, max} = 966 \pm 1.5$ K, while the excitation of the symmetric one is rather moderate with $T_{v12, max} = 438.4 \pm 1.2$ K. This confirms the preferential excitation of the asymmetric stretch mode of $CO_2$, as reported with different gas mixtures in experiments [17, 27] and simulations [6, 7]. In addition, the elevation in the gas temperature is even more insignificant with a peak value of only $334.6 \pm 0.6$ K. This low gas temperature is shown to be very desirable for $CO_2$ dissociation since it helps to minimize both the recombination process (CO + O + M → $CO_2$ + M) and the deexcitation of the higher vibrational states, which may enhance the most energy-efficient pathway of $CO_2$ dissociation (i.e. through the vibrational levels), through VT relaxation considering the strong temperature dependence of both processes[5, 47]. After $t = 3$ μs, all temperatures start to quasi-exponentially decrease towards an equilibrium value $T_{eq} = 311.1 \pm 0.2$ K before the next discharge pulse. Experimental results also show that this equilibrium temperature increases slightly (~10 K) with an increase of pulse length (from 150 ns to 200 ns) and applied voltage (from 3 kV to 3.5 kV) due to the increase in the pulse energy. Furthermore, it also can be seen that the symmetric stretching (and bending) modes present a faster decay than the asymmetric mode, due to the generally larger V-V and V-T relaxation cross sections of the symmetric bending modes compared with the asymmetric one [10, 48].

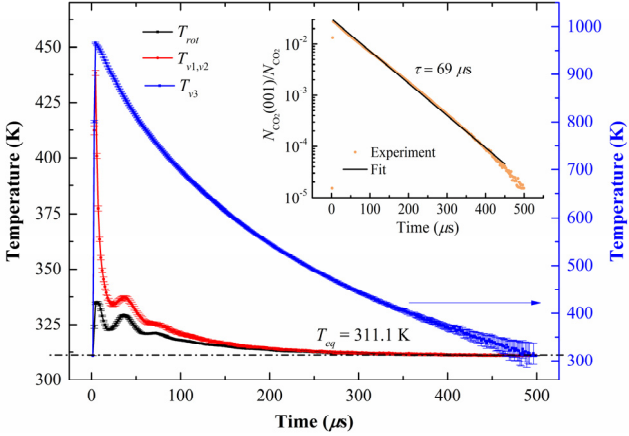

Figure 9. Time evolution of the best-fit rotational temperature and vibrational temperatures in the afterglow of the ns-discharge with 10% $CO_2$ + He. The discharge condition is the same as that in Fig. 7. The inset figure depicts the evolution of the normalized number density (001) state versus time together with the best-fitted exponential decay.

*4.3 Deexcitation of the asymmetric mode and acoustic wave perturbation*

The inset in Fig. 9 depicts the time evolution of the number density of the $CO_2$ asymmetry mode (001) state, which essentially determines the vibrational temperature, $T_{v3}$, in the broadband spectrum fitting. As can be seen, the number density of $CO_2$ (001) state follows very well an exponential decay with a time constant of $\tau = 69$ $\mu$s. The measured decay constant matches well with the theoretical values $\tau = (\tau_{VT}^{-1} + \tau_{Wall}^{-1})^{-1} = 73$ $\mu$s by considering its loss mechanisms of VT transfer with He and the deexcitation on the wall ($\tau_{VT} \sim 127$ $\mu$s estimated with $\sigma \sim 2*10^{-20}$ cm$^2$ @ 300K [48], $\tau_{Wall} \sim 167$ $\mu$s estimated with destruction probability $\gamma = 0.2$ with Pyrex surface [10]). This suggests that the deexcitation of (001) state in our discharge condition is mainly dominated by the VT transfer with the buffer gas and the deexcitation on the wall. This also explains the relatively low dissociation degree of $CO_2$ in our working condition since the nearly resonant collisional up-pumping VV process [49], which is believed to be important in the plasma-assisted $CO_2$ dissociation by producing higher asymmetric vibrational states for dissociation, does not contribute much at our conditions. In addition, it should be mentioned that although the vibrational temperature $T_{v3}$ is mainly inferred from the number density of the first vibrational level (001), both experiments and state-to-state simulation [6, 7] have shown a nearly-Boltzmann distribution among the asymmetric mode ($v_3 \leq 5$), which makes the measured (001) state a good representative of the asymmetric vibrational mode.

In addition to the slow decay of $T_{v3}$ with respect to $T_{rot}$ and $T_{v1,2}$, another pronounced feature of the temporally resolved temperatures is the oscillation in the best-fit $T_{rot}$ and $T_{v1,2}$ within the first 100 $\mu$s. A similar change is also observed in the best-fit total pressure, as shown in Fig. 10(b) plotted on a logarithm scale. Before further discussing the potential origin of this oscillation, the results for the gas temperature and pressure from the broadband spectrum fitting are further validated by the conventional two-line thermometry in TDLAS[29]. It is implemented by fitting the temporally resolved absorption peaks "0" and "3" with a Voigt profile. Since both peaks belong to the same vibrational state (000), the fitted area ratio of these two peaks depends only on the rotational temperature. The Lorentzian broadening components of both peaks are proportional to the total pressure as shown in Eq. (5) if the slight temperature change is ignored. The area ratio and the collisional broadening from the single peak fitting are shown in Fig. 10, together with the rotational temperature and pressure results referred from the broadband spectrum fitting. Several conclusions can be obtained: First, the overall ratio of the best-fit Lorentzian broadening of peak "0" and "3" in Fig. 10 (b) is 1.18, which corresponds well with the theoretical ratio of broadening coefficients of these two peaks, suggesting the reliability of the single peak fitting algorithm. Secondly, the perfect agreements between the time-resolved area ratio and rotational temperate and between the Lorentzian broadenings and total pressure validate the accuracy of the broadband spectrum fitting and the existence of such oscillation.

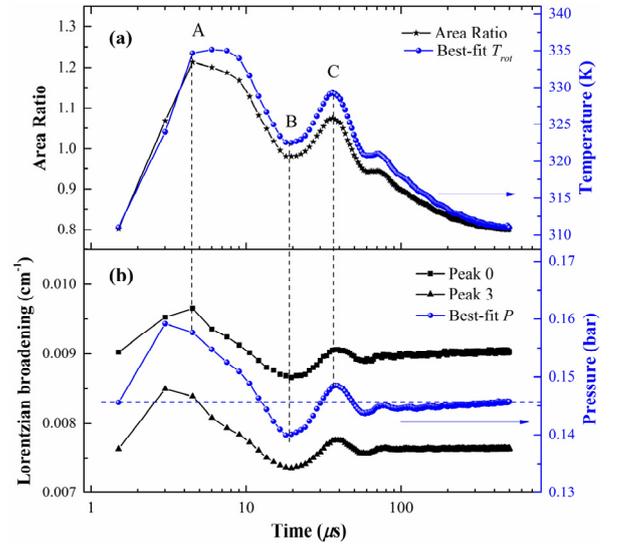

Figure 10. Validation of the best-fit rotational temperature and pressure from the broadband spectrum fitting with the generally used two-line thermometry method (with peaks "0" and "3") in conventional TDLAS.

To further look into this temperature and pressure oscillation process, we estimated the specific heat capacity of the gas mixture, i.e. ~10% $CO_2$ + 90 % He ignoring the slight amount of dissociation products CO or $O_2$, to be $\gamma_{mix}$ = 1.583 ($\gamma_{CO2}$ = 1.275, $\gamma_{He}$ = 5/3). It should be noted that the specific heat capacity of $CO_2$, $\gamma_{CO2}$, used here is calculated at a thermalized averaged gas temperature $T$ = 330 K. ($C_p$ = A + B*$t$ + C*$t^2$ + D*$t^3$ + E/$t^2$, $t$ = $T$/1000, with constants A, B, C, E adopted from [50]). Based on $\gamma_{mix}$, the pressure changes in A→B and B→C (Fig. 10 (a)) can be estimated to be 15.2 and 7.9 mbar, respectively, assuming an adiabatic process ($P^{1-\gamma}T^\gamma$ = const.). In fact, the observed pressure changes in the experiments, 17.4 and 8.6 mbar, correspond well with the theoretical estimations. The small overshoot in the experiments might be attributed to the underestimation of $\gamma_{mix}$ caused by a slightly small $C_p$ for $CO_2$, since the asymmetric mode of $CO_2$ in our experiment is apparently un-frozen, which yields extra degrees of freedom for heat absorption.

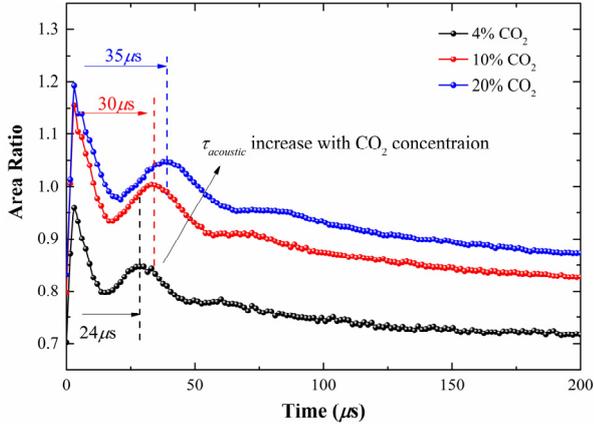

Figure 11. The area ratio of peak "3" and "0" deduced from the single-peak fitting with different $CO_2$ concentrations.

Furthermore, as shown in Fig. 10, the oscillation of the gas pressure shows a damped oscillation with a period around 30 $\mu$s, which gives us some hints about its origin. One potential reason might be the acoustic wave produced by the pressure perturbation, i.e. the pressure perturbation produced by the discharge propagates with sound speed to the edge of the discharge reactor and finally forms a standing wave in the discharge gap. Since the laser absorption spectroscopy is a line-of-slight technique, the pressure we measured is in fact the averaged pressure along the electrodes. By considering the sound speed in the gas mixture and the geometry of our discharge, the period of the fundamental mode standing wave is $\tau_{acoustic}$ = $\lambda$/$c_{mix}$ = 2 × $l_{charc}$/$c_{mix}$ ~ 28.5 $\mu$s, where $l_{charc}$ is the characteristic length between two open ends, i.e. the distance between the gas inlet and the discharge boundary, and $c_{mix}$ ~ 700 m/s ($c_{mix}$ is the sound speed in 10% $CO_2$/He gas mixture. As can be seen, this acoustic time scale agrees very well with the observed period of the pressure oscillation. To further validate this hypothesis, the pressure oscillation was measured with different $CO_2$/He concentrations as shown in Fig. 11. A clear increase of the oscillation period is observed with increasing $CO_2$ concentration due to the decrease of sound speed in the gas mixture. The measured oscillation periods, 24, 30, and 35 $\mu$s correspond well with the theoretical values, 23.5, 28.5, and 34.2$\mu$s, which supports the origin of the pressure oscillation and also demonstrates the high sensitivity of this measurement method.

## 5. Conclusion

In this paper, we have presented a method to determine the time evolution of $CO_2$ ro-vibrational excitation with the compact quantum cascade laser absorption spectroscopy. With a well-selected wavelength window around 2289.0 cm$^{-1}$ and a single QCL operating at continuous mode, the rotational temperature, vibrational temperatures for both symmetric and asymmetric modes, as well as the $CO_2$ density can be simultaneously determined with both high accuracy and temporal resolution.

The proposed method is further applied to monitor the rotational and vibrational temperatures of $CO_2$ in the afterglow of a nanosecond discharge with a temporal resolution of 1.5 $\mu$s. The discharge operates with 10% $CO_2$ + He at 145 mbar with a frequency of 2 kHz and a pulse length of 150 ns. Experimental results show that both gas temperature and vibrational temperatures keep increasing in the early afterglow and reach the corresponding maximum value at 3 $\mu$s. The non-thermal feature and the preferential excitation of the asymmetric stretch mode of $CO_2$ were clearly observed, with a peak vibrational temperature of $T_{v3, max}$ = 966 ± 1.5 K, $T_{v12, max}$ = 438.4 ± 1.2 K and $T_{rot}$ = 334.6 ± 0.6 K. This moderate elevation of gas temperature is shown to be beneficial to the $CO_2$ dissociation since it helps to minimize the reverse recombination process and the deexcitation of the higher vibrational states. Furthermore, in the relaxation process, the measured number density of $CO_2$ (001) state presents an exponential decay with a time constant of $\tau$ = 70 $\mu$s, which is mainly attributed to the VT transfer with He and the deexcitation on the wall. Moreover, within the first 100 $\mu$s of the relaxation process, a synchronous oscillation of gas temperature and total pressure was observed with a period of 30 $\mu$s. It is very likely to be caused by the acoustic wave according to the similar acoustic time scale $\tau_{acoustic}$ in our experimental condition and its dependence on the gas components.

Based on these preliminary results, further effort will be made to look into the excitation of $CO_2$ in the active discharge phase (within the discharge pulse) in combination with a faster detector. Also, the time-dependent measurement of the higher vibrational state in the asymmetric mode is also suggested to further check the effect of the vibrational ladder-climbing pathway on the plasma-assisted $CO_2$ dissociation.

**Acknowledgments**

This work is within the DFG funded SFB1316 Project "Transient atmospheric plasmas: from plasmas to liquids to solids". Yanjun Du acknowledges the financial support from the Alexander von Humboldt Foundation. The authors are indebted to the valuable discussions with Professor Achim von Keudell on the non-equilibrium $CO_2$ spectrum calculation.